\documentstyle[graphicx,aps,amsmath]{revtex}

\newcommand{\intii}{\int_{-\infty}^\infty}
\newcommand{\hpsi}{\hat{\psi}}
\newcommand{\hpsid}{\hat{\psi}^\dag}
\newcommand{\hPsi}{\hat{\Psi}}

\newcommand{\hphi}{\hat{\phi}}

\newcommand{\hagp}{\hat{a}_{\text{\tiny GP}}}
\newcommand{\hagpd}{\hat{a}_{\text{\tiny GP}}^\dag}
\newcommand{\ha}{\hat{a}}
\newcommand{\had}{\hat{a}^\dag}
\newcommand{\hb}{\hat{b}}
\newcommand{\hbd}{\hat{b}^\dag}
\newcommand{\hdpsi}{\hat{\delta \psi}}
\newcommand{\hdpsid}{\hat{\delta \psi}^\dag}

\newcommand{\al}{\alpha}
\newcommand{\als}{\alpha^*}
\newcommand{\alc}{\alpha^+}
\newcommand{\hH}{\hat{H}}
\newcommand{\hO}{\hat{O}}
\newcommand{\hOd}{\hat{O}^\dag}
\newcommand{\hX}{\hat{X}}
\newcommand{\bx}{{\mathbf{x}}}
\newcommand{\bB}{{\mathbf{B}}}
\newcommand{\calK}{{\cal K}}
\newcommand{\calL}{{\cal L}}
\newcommand{\dx}{\mbox{d}x \,}
\newcommand{\psigp}{\psi_{\text{\tiny GP}}}
\newcommand{\psigpbar}{\overline{\psi}_{\text{\tiny GP}}}

\newcommand{\ket}[1]{\left|#1\right\rangle}
\newcommand{\bra}[1]{\left\langle #1\right|}
\newcommand{\braket}[2]{\left \langle #1 \right|\left. #2 \right \rangle}
\newcommand{\expect}[1]{\left\langle #1\right\rangle}
\newcommand{\smexpect}[1]{\langle #1\rangle}
\newcommand{\ngp}{\hat{n}_{\text{\tiny GP}}}
\newcommand{\sech}{\mathop{\mathrm{sech}}}

\begin{document}
\preprint{mikes}
\title{Dynamical quantum noise in trapped Bose-Einstein condensates}
\author{$\mbox{M. J. Steel}^{1,2}$, 
$\mbox{M. K. Olsen}^1\footnote{Current address:
Laborat\'{o}rio de \'{O}tica Qu\^{a}ntica,
Instituto de F\'{\i}sica, Universidade de S\~{a}o Paulo - USP,
CX Postal 66318, S\~{a}o Paulo - SP,
05389 - 970, Brazil} $, 
$\mbox{L. I. Plimak}^1$, 
$\mbox{P. D. Drummond}^{1,3,4}$,
$\mbox{S. M. Tan}^1$,
$\mbox{M. J. Collett}^1$,
$\mbox{D. F. Walls}^1$ and 
$\mbox{R. Graham}^{1,5}$}

\address{1. Department of Physics, University of Auckland, Private Bag 92 019,
Auckland, New Zealand. \\
2. School of Mathematics, Physics, Computing and Electronics,
        Macquarie University, North Ryde, New South Wales 2109, Australia. \\
3. Institute for Theoretical Physics, University of California,
Santa Barbara, California 93106--4030. \\
4. Department of Physics, University of Queensland, St Lucia, 
	Queensland 4072, Australia. \\
5. Universit\"{a}t GH Essen, Fachbereich Physik, D45117, Essen,Germany.}

\date{\today}
\maketitle

\begin{abstract}
  We introduce the study of dynamical quantum noise in Bose-Einstein
  condensates through numerical simulation of stochastic partial
  differential equations obtained using phase space representations.
  We derive evolution equations for a single trapped condensate in
  both the positive-$P$ and Wigner representations, and perform
  simulations to compare the predictions of the two methods. The
  positive-$P$ approach is found to be highly susceptible to the
  stability problems that have been observed in other strongly
  nonlinear, weakly damped systems. Using the Wigner representation,
  we examine the evolution of several quantities of interest 
  using from a variety of choices of initial
  state for the condensate, and compare results to those for single-mode 
  models.
\end{abstract}

\section{Introduction}
A key focus of the explosion of interest in the new dilute
atomic gas Bose-Einstein condensates~\cite{and95,bra97,dav95} has been
the study of the time-evolution of condensates from some initial
state. 
Amongst many works, there have been theoretical investigations
of the way condensates react to a range of perturbations,  
such as
``shaking'' the trap to excite sound waves~\cite{and97a,kav97,zar98},
removing a potential barrier to allow two condensates to
interfere~\cite{and97,sco98}, applying electromagnetic fields to
transfer condensate population into other possibly untrapped
states~\cite{bal97,jac98,mew97,mat98}, or ``stirring'' a condensate to
excite vortices~\cite{jac97}. All but the last of these
effects have already been demonstrated experimentally.  A common
element in the theoretical works on these topics is the description of
the condensate using the time-dependent Gross-Pitaevskii equation
(GPE) or coupled GPEs (or their approximate hydrodynamic versions). 
{The GPE can be derived as an equation for the condensate 
amplitude assuming the condensate state is a multi-mode 
coherent state (on the concept of coherent states
see~\cite{crispin2,wal94}). Hence an implicit assumption underlying 
these approaches is that the condensate is adequately
described as a coherent state. However, a number of experiments are
now exploring issues such as coherence~\cite{bur97,nistlaser} and
the diffusion of relative phase between two condensates~\cite{hal98}. These
concepts are familiar from optical systems, but additional factors
arise in condensate physics such as the dispersion associated with 
the nonzero atomic mass, and especially the effects of atomic interactions. 
In particular, several early models for atom 
lasers~\cite{qua96,wis96,hol96a} suggest that coherence properties may be 
strongly influenced by the nonlinear interactions.
Moreover, one of the principal themes
of quantum optics is the idea that processing of quantum noise
by nonlinearities leads to interesting statistical
properties~\cite{crispin2,wal94}.  It thus becomes important to
consider the nature of the condensate beyond the
coherent state, and in particular the influence of quantum noise on
the coherence of the condensate and any eventual ``atom laser''.  For
example, we may ask how different loss profiles for an output
coupler~\cite{kne98} might affect the noise statistics of an atom
laser?

In fact, there have been a number of studies into
intrinsically quantum dynamical effects, in particular the collapse and revival
of the relative phase between two coupled
condensates~\cite{mil97,wri97} and the robustness of such effects
against environmental decoherence~\cite{ste97,ruo98}.
However, these studies have been restricted to one or two modes and
assume the condensate wave function is independent of the number of
atoms. For large condensates, these approximations are not necessarily valid
and a many-mode approach is required.
A single mode model may give an estimate for the
phase diffusion time~\cite{wri96} for example, but can never describe
spatial coherence properties or the role of local density and phase
fluctuations. Hence, there is a need for techniques to treat the 
quantum dynamics of the condensate using a fully spatially dependent field 
rather than a few mode approach.

Closely related issues are well-known in the field of quantum optical solitons
and nonlinear quantum optics in general. In that situation, the propagation
of the optical field is governed by a quantum nonlinear Schr\"{o}dinger 
equation that includes the effects of fiber dispersion and the Kerr
nonlinearity of the medium~\cite{car87,dru87}. Such a model leads 
to the prediction that a soliton pulse injected into the fiber experiences
squeezing (for general discussions of squeezing see~\cite{crispin2,wal94},)  
in both the electric field 
amplitude~\cite{car87,dru87} and the photon number~\cite{fri96,wer96}. 
Both types of squeezing have now been observed~\cite{she86,spa97}. 
While estimates for the squeezing have also been
obtained from single mode models of the Kerr
nonlinearity~\cite{wal94}, the presence of fiber dispersion means that accurate
results can only be obtained from a multi-mode treatment of the
full quantum field. The Heisenberg field operator describing a 
BEC confined in a one-dimensional
potential well obeys a strikingly similar equation to that of the fiber
soliton system, differing only by the addition of the trapping potential 
and the interpretation of the dispersive term which in a condensate
represents the kinetic energy. Thus given that 
multi-mode models have proved essential for the accurate prediction of
quantum soliton properties, it is reasonable to assume the same may hold
true in Bose condensates. In fact, we see below that the nonlinearity occurring
in the condensate problem is typically far larger than for the soliton case,
and thus the role of quantum noise should be more important.

While we thus have strong motives for seeking the complete evolution of the
field operator, such a calculation is at first sight a formidable task,
if for no other reason than that the Hilbert space for the system is truly
vast. The numerical calculation of the evolution of just a single operator 
with significant excitation requires a large basis. The problem of the
field is far worse.
Nevertheless, in this paper we intend to demonstrate how techniques of
quantum optics may be used to provide a complete description of the
condensate field operator such that we can calculate virtually any desired 
quantum expectation value. The key to our approach is the representation
of the density operator using phase-space quasi-probability functions.
These functions then lead naturally to a description in terms of
\emph{classical} fields which are subject to evolution equations
similar to the semi-classical time-dependent Gross Pitaevskii equation~(GPE) 
satisfied by the mean field but with the crucial addition of stochastic 
driving terms. These terms do not correspond to any physical 
noise sources, but are defined in such a way as to recapture exactly
the operator character of the fully quantum-mechanical field.
In particular we use two representations: the positive-$P$ function 
and the well-known Wigner distribution.
While we can perform exact calculations in the
positive-$P$ representation, we find the system rapidly succumbs to
the instabilities that have been observed previously for that 
representation~\cite{smi89,gil97}. Therefore we also consider an approximate
but robust method using the Wigner representation. We are then able to
extract a large range of interesting averages.

It is worth noting that in
terms more familiar in conventional quantum field theory, 
the stochastic techniques we present in this paper constitute a  
method of  numerically 
evaluating path-integral representations of quantum 
field averages (see Plimak {\em et al\/}~\cite{EPL,JPA1-2,JPA3}). 
The phase-space of the classical fields 
is in these terms the space of the Feynman paths, 
while the phase-space 
quasi-probability functions are measures over 
the respective path integrals.   
These measures are constructively characterised by  
the corresponding stochastic evolution equations, 
which allows the path integrals to be calculated.  
This point highlights the significance of the positive-$P$ representation.
Providing certain boundary conditions are satisfied~\cite{crispin2,wal94},
the positive-$P$ is an \emph{exact} method for propagating the field
in real (as opposed to imaginary,) time. 
While it may blow up in certain cases as time progresses, 
all other exact methods fail at all times!
Direct integration of the Feynman path integrals for example, is 
numerically useless due to the oscillatory phase factors.

The  stochastic techniques  are well-known  in quantum optics,
and may be familiar to readers with a 
background in that subject. However, many of these ideas may be new
to those who have come to Bose-Einstein condensation from other
disciplines. Therefore we have taken a pedagogic approach in deriving the
fundamental stochastic equations. Before treating the full quantum field
problem, we review the techniques in the single-mode approximation for which 
the Hamiltonian corresponds to the anharmonic oscillator. The analysis for
the complete field then follows in a natural way. Readers already 
well-versed in the stochastic approach to quantum dynamics may wish 
to pass over this earlier background in section~\ref{sec:one-mode-prob}.

The paper is structured as follows. In
section~\ref{sec:Techn-Prop-Quant} we provide a detailed demonstration
of the techniques for propagation of quantum fields using phase space
representations. After stating the complete problem in
section~\ref{sec:Trapp-Bose-Einst}, we simplify to the corresponding one-mode
Hamiltonian in section~\ref{sec:one-mode-prob} 
and derive the equivalent stochastic equations in the
positive-$P$ representation in detail.  We generalize this approach to
the full field in section~\ref{sec:multi-mode-prob}, and find an
approximate but more stable method using the Wigner function in
section~\ref{sec:Trunc-Wign-repr}.  
There is considerable freedom in the choice of initial states
for our simulations.  We discuss these issues and present some natural
choices in section~\ref{sec:Initial-States}. Our numerical results
illustrating some of the possibilities of the stochastic approach are
given in a range of section~\ref{sec:Results} before we conclude.

\section{Techniques for Propagation of Quantum Fields}
\label{sec:Techn-Prop-Quant}
\subsection{General Ideas}
There are a number of well-established techniques in quantum optics
for the propagation of a complete quantum field.  Typically, these
ideas involve a generalization of standard procedures for finding the
time evolution of averages in a system with a single mode or a few
modes~\cite{crispin2,wal94}. In summary, the procedure is as follows. 
The system density operator is expressed
in the coherent state basis using a quasi-probability function such as
the $P$, Wigner or positive-$P$ distributions. The master equation
describing the evolution of the density operator is converted to an
equivalent partial differential equation (PDE) for the distribution.
If certain conditions are satisfied~\cite{crispin2,wal94}, the PDE may
then be converted to a set of classical stochastic ordinary
differential equations (Langevin equations)
which yield quantum expectation values as
ensemble averages of moments of the phase space variables.  This
procedure has the significant advantage of providing a natural
numerical implementation in which we calculate the evolution of a
small number of phase space variables rather than that of a
very large number of variables describing the density matrix or a
distribution function on a large complex grid. This advantage becomes
essential for systems of several modes for which the Hilbert space is
so large that a direct numerical simulation would be impossible.
Now when we consider a quantum field, unless we are
fortunate enough to have an analytic solution, the problem must be
treated numerically as a system with a large but finite number of
modes and the associated Hilbert space is truly vast. A thousand atoms
with just one hundred modes for example, occupy a Hilbert space
of dimension $100^1000$. The stochastic
treatment is now vital.  One uses essentially the same procedure but
works with \emph{functional} PDE's, and thus obtains stochastic
equations for a classical field~\cite{car87,dru87,grahak70,ken88}. 
In the following sections we provide a detailed
derivation for the case of a Bose condensate in one dimension.

\subsection{Trapped Bose-Einstein condensates}
\label{sec:Trapp-Bose-Einst}
We model a one-dimensional system by assuming a highly anisotropic
harmonic trap with the longitudinal and radial trap frequencies
($\omega_z$ and $\omega_r$ respectively ) satisfying
$\lambda=\omega_z/ \omega_r \ll 1$. Below, we use parameters
corresponding to a cigar-shaped trap such as that in
Refs.~\cite{and97,mew97}. With strong radial confinement, we assume
the nonlinearity plays a negligible role in the radial direction. The
field operator is then assumed to factorize with its transverse
dependence completely described by a coherent state occupation of the
lowest mode of the trap. So the Heisenberg picture boson
field operator has the form
\begin{equation}
  \label{eq:fieldop}
  \hPsi(\bx) = \left(\frac{m\omega_r}{\pi\hbar}\right)^{1/2}\exp\left(-\frac{m\omega_r r^2}{2\hbar}\right)\hphi(z,t).
\end{equation}
Adopting harmonic oscillator units in the axial direction with
$a_0=\sqrt{\hbar/(m\omega_z)}$, $\tau = \omega_z t$, $x=z/ a_0$ and
$\hpsi(x,\tau) = \sqrt{a_0}\hphi(z,t)$, the one-dimensional 
second-quantized Hamiltonian is
\begin{equation}
  \label{eq:hamil}
  \hH = \int_{-\infty}^{\infty} \dx \hpsid \calK \hpsi 
  + \frac{\Gamma}{2} \int_{-\infty}^{\infty} \dx \hpsid\hpsid\hpsi\hpsi,
\end{equation}
where $\calK$ is the linear operator
\begin{equation}
  \calK= -\frac{1}{2}\frac{\mbox{d}}{\dx^2}+\frac{1}{2}x^2-\mu , 
\end{equation}
$\mu$ is the scaled chemical potential, and $\Gamma = 2a/(\lambda a_0)$ is the
scaled nonlinear constant with $a$ the s-wave scattering length. 
Our ultimate aim is to calculate (multi-time) averages of $\hpsi$ under
the evolution induced by the Hamiltonian~(\ref{eq:hamil}). More generally, 
the system may include damping in the form of couplings to
atom reservoirs. In this case the system is described by a density
matrix satisfying a master equation
\begin{equation}
\label{eq:master}
\frac{d\rho}{d\tau} = \calL \rho
\end{equation}
where the Liouvillian $\calL$ is a superoperator that acts to the right 
in the fashion
\begin{equation}
\calL \rho = -i [\hH,\rho]+
\sum_j \frac{\kappa_j}{2}(2\hO_j\rho\hOd_j- \hOd_j\hO_j\rho-\rho\hOd_j\hO_j),
\end{equation}
and the $\hO_j$ are operators describing the bath couplings with strengths
$\kappa_j$.
While we do not include damping in the present work, 
it is convenient to work in  a density matrix formalism.
We now develop stochastic descriptions of the dynamics
in both the $P$ and Wigner
representations, explaining the method for the $P$ representation in
detail.

\subsection{One mode problem}
\label{sec:one-mode-prob}
\subsubsection{$P$ representation}
\label{sec:P-rep}
To illustrate the ideas underlying a phase-space approach without
the notational baggage of the full many-mode problem, we begin by treating the
single mode limit of Eq.~(\ref{eq:hamil}). The atomic field
is assumed to be described by a single mode operator $\ha(t)$ with
an associated mode function $\psigp$ determined by the solution to 
the time-independent GPE. The single mode Hamiltonian is
\begin{equation}
\hH^{(1)} = \bar{\omega} \had\ha + \frac{\chi}{2} \had\had\ha\ha,
\end{equation}
with $\bar{\omega} = \intii \dx \psigp^* \calK\psigp$
and $\chi=\Gamma \intii \dx |\psigp|^4$.
The first step in our attempt to obtain a stochastic description is 
to express the density matrix in a diagonal coherent state basis using 
the Glauber-Sudarshan~$P$ function~\cite{crispin2,wal94}:
\begin{equation}
  \label{eq:gsP}
  \rho = \int d^2\al\, |\al\rangle\langle\al| P(\al),
\end{equation}
where $|\al\rangle$ is a coherent state with the c-number 
complex amplitude $\alpha $.
It is tempting to consider $P(\al)$ as a probability distribution for
the density matrix over the coherent states. However, while $P(\al)$ is real,
for non-classical states it may be highly singular and/or take on negative 
values~\cite{crispin2,wal94}. 
It is proved in Ref.~\cite{crispin2} that there is a unique $P$ 
function for every density matrix. The quantum averages of interest are 
found as  moments of the $P$ distribution which correspond to
\emph{normally}-ordered expectation values:
\begin{equation}
  \label{eq:Pmom}
  \expect{\ha^{\dag m}\ha^n}=\mbox{Tr}\{\ha^{\dag m}\ha^n\rho\}=
 \int d^2\al\,\al^{*m}\al^nP(\al),
\end{equation}
for integer $m,n\ge0$. Arbitrarily-ordered averages can always be
found by first  rewriting them in terms of normally-ordered quantities.

We now need an equation for the time-evolution of the $P$ function.
Using $|\al\rangle = 
\exp(-|\al|^2/2)\exp(\al\had)\exp(-\alc\ha)\ket{0}$ and 
the definition of the $P$ function it is not hard to demonstrate the 
operator correspondences:
\begin{align}
  \label{eq:Pequiv}
  \ha\rho &\leftrightarrow \al P(\al) , &
  &\had \rho \leftrightarrow 
     \left(\alc-\frac{\partial}{\partial\al }\right) P(\al) 
 \nonumber \\
  \rho \ha &\leftrightarrow 
     \left(\al-\frac{\partial}{\partial\alc }\right) P(\al), &
  &\rho \had \leftrightarrow \alc P(\al) .
\end{align}
We have introduced the unusual notation $\alc$, which for the moment is
to be read as the ordinary complex conjugate $\als$.
Substituting these correspondences in the master equation 
$\dot{\rho}=-i[\hH^{(1)},\rho]$ we obtain a 
Fokker-Planck equation for the time-evolution of $P(\al)$
\begin{equation}
  \label{eq:Pfok}
  \frac{\partial P}{\partial \tau} =
 \left\{-\frac{\partial}{\partial\al}[-i(\bar{\omega}\al+\chi\alc\al^2)]
       -\frac{\partial}{\partial\alc}[i(\bar{\omega}\alc+\chi\al\al^{+2})]
+\frac{1}{2}\frac{\partial^2}{\partial \al^2}(-i\chi\al^2)
+\frac{1}{2}\frac{\partial^2}{\partial \al^{+2}}(i\chi\al^{+2})\right\}P(\al,\alc).
\end{equation}
Note that despite the appearance of this equation, 
$\al$ and $\alc$ are not to be treated as independent variables
as they are complex conjugates~\cite{crispin2,wal94}.

Equation~(\ref{eq:Pfok}) is exact and completely equivalent to the master 
equation. Our motivation in obtaining it is based on the fact that any
Fokker-Planck equation with a  positive-definite diffusion matrix may be 
exactly rewritten in the language of stochastic differential or Langevin 
equations~\cite{crispin1}.
To be precise, consider a Fokker-Planck equation of the form
\begin{equation}
  \label{eq:fokplanck}
  \frac{\partial F}{\partial t} = -\sum_j \frac{\partial}{\partial x_j}
A_j(\bx,t)P+\frac{1}{2}\sum_{jk}\frac{\partial}{\partial x_j}
\frac{\partial}{\partial x_k} \left[\bB(\bx,t)\bB^T(\bx,t)\right]_{jk}F,
\end{equation}
in which the diffusion matrix $D=\bB\bB^T$ is clearly positive definite.
Then a third equivalent description is given by the system of
stochastic equations
\begin{equation}
  \label{eq:stoch}
  \frac{d\bx}{dt}=A(\bx,t)+\bB(\bx,t){\mathbf{E}}(t),
\end{equation}
where the real noise sources $E_j(t)$ have zero mean and satisfy
$\overline{E_j(t)E_k(t')}=\delta_{jk}\delta(t-t')$. These equations
(and all other stochastic equations in this paper) are to be
interpreted in the Ito approach to stochastic calculus. 
A complete discussion of the techniques of
stochastic calculus and the connection between the Fokker-Planck and
Langevin descriptions is provided by Gardiner~\cite{crispin1}.
By making such a transformation we would apparently have achieved our aim 
of a stochastic description of the quantum dynamics, and
could calculate expectation values by taking 
ensemble averages of moments of the phase space variables $x_j$.  

\subsubsection{Positive-$P$ representation}
\label{sec:pos-P-rep}
Unfortunately, the diffusion matrix in Eq.~(\ref{eq:Pfok}) is clearly not 
positive-definite and the preceding equivalence does not apply.
However, Drummond and Gardiner~\cite{dru80a} have shown that in such cases, 
the situation may be rescued by introducing the ``Positive-$P$'' function
which represents the density matrix as an integral over two
independent variables 
\begin{equation}
\rho = \int d^2\alpha d^2\beta \, \frac{\ket{\alpha}\bra{\beta^*}}
        {\braket{\beta^*}{\alpha}} P(\alpha,\beta).
\end{equation}
It can be shown that with this definition the positive-$P$ function is
analytic and can be chosen positive for any density
matrix~\cite{crispin2,dru80a}. The crucial step comes here.  Referring
to Eq.~(\ref{eq:Pfok}), we now consider $\al$ and $\alc$ as
\emph{independent} quantities and by making the identification
$\beta=\alc$, we may read Eq.~(\ref{eq:Pfok}) as the Fokker-Planck
equation in the positive-$P$ representation corresponding to the
original master equation. Moreover, by writing $P(\alpha,\beta)$
as a function of four real variables rather than
two complex variables, one may show that the resulting $4\times 4$ diffusion
matrix is always positive-definite~\cite{crispin2,dru80a}. Thus in the
positive-$P$ representation, it is always possible to derive an
equivalent Langevin equation description using Eq.~(\ref{eq:stoch}).
In the present case, Eq.~(\ref{eq:Pfok}) leads to the stochastic
system
\begin{align}
\label{eq:onestoch}
i\frac{d \al}{d \tau} &= (\bar{\omega}\al+\chi\alc\al^2) + \sqrt{i\chi}\al\eta_1(t) , \nonumber \\
i\frac{d \alc}{d \tau} &= -(\bar{\omega}\alc+\chi\al\al^{+2}) + \sqrt{-i\chi}\alc\eta_2(t),
\end{align}
where $\eta_i(t)$ for $i=1,2$ are delta-correlated in time with zero-mean.
Note that $\al$ and $\alc$ experience different noise sources, so that
even if they are  conjugate  at $\tau=0$ they do not remain so. 

We emphasize that Eqs.~(\ref{eq:onestoch}) are completely equivalent
to the master equation. Any expectation value that can be found from
the density operator may equally be found by ensemble averaging over many
trajectories using the stochastic equations. Being c-number variables,
$\al$ and $\alc$ do not satisfy the commutation properties of the
operators $\ha$ and $\had$. Nevertheless, through the inclusion of the
noise sources $\eta_i$ and the insistence on normal ordering when
taking averages, they still account for the complete quantum dynamics.
This equivalence of the stochastic and operator approaches has been
demonstrated explicitly in the context of optical fiber solitons in a
recent paper of Fini~\emph{et al}~\cite{fin98}.  It is important
to note that the
$\eta_i$ do \emph{not} correspond to any physical noise sources, but
are included only to recapture the commutation relations of the
operators. In this sense they are quite
distinct from the operator valued noise sources that appear in 
``quantum Langevin equations''~\cite{crispin2,wal94}.  
In fact in the positive-$P$ representation,
Plimak~\emph{et al}~\cite{JPA3}
 have pointed out that there is some freedom in the
precise form of the noise terms, which can be exploited to improve
convergence properties dramatically.  
We also mention a well-known difficulty with
the positive-$P$ representation.  The independence of the noise
sources driving $\al$ and $\alc$ can in some cases lead to wild
trajectories that prevent convergence of the ensemble
averages~\cite{smi89,gil97,wer97}. This is indeed true in the
present case and we consider this problem in some
detail in later sections.  As the properties of the anharmonic
oscillator are well-known, we do not present simulations of the
one-mode equations~(\ref{eq:onestoch}), but proceed directly to the
multi-mode field problem.

\subsection{Multi-mode problem}
\label{sec:multi-mode-prob}
By analogy with the single-mode problem, in which single-mode operators
were replaced by classical variables driven by white noise sources,
we might expect that a complete quantum field can be 
replaced by classical fields suffering independent noise sources at every
point in space. There are a number of ways of proving this claim. A
straight-forward (if notationally cumbersome,) method is to expand the field 
in a complete set of modes and mode operators and proceed by a 
direct generalization of the method in 
the previous section~\cite{ken88}. However,
a more concise derivation is obtained by introducing 
the functional $P$-distribution ~\cite{grahak70} 
\begin{equation}
  \label{eq:Pdef}
  P(\{\psi,\psi^*\},\tau)=\left.\rho^{(a)}(\{\hpsi,\hpsid\},\tau)
  \right |_{
    \begin{array}{l}
      \scriptstyle \hpsi\rightarrow \psi \\  
      \scriptstyle \hpsid\rightarrow \psi^*
    \end{array}}
\end{equation}
where $\rho^{(a)}$ denotes the density operator $\rho(\tau)$
antinormally ordered with respect to the field-operators $\hpsi,
\hpsid$ in the Schr\"{o}dinger picture. Putting the master equation 
obtained from the Hamiltonian~(\ref{eq:hamil}) into antinormal order,
and  using the following 
functional analogues of the operator correspondences~(\ref{eq:Pequiv}):
\begin{align}
   \hpsi\rho &\leftrightarrow \psi P(\psi) , 
  & &\hpsid \rho \leftrightarrow 
     \left(\psi^{+}-\frac{\delta}{\delta\psi }\right) P(\psi) ,
 \nonumber \\
  \rho \hpsi &\leftrightarrow 
     \left(\psi-\frac{\delta}{\delta\psi^{+} }\right) P(\psi),
  & &\rho \hpsid \leftrightarrow \psi^{+} P(\psi) .
\end{align}
one finds the functional Fokker-Planck equation,
\begin{equation}
  \label{eq:fokker}
  \frac{\partial P}{\partial \tau}=\intii \dx \left\{-\frac{\delta}{\delta
    \psi(x)}\left[
  -i\left(\calK\psi(x) +\Gamma |\psi(x)|^2\psi(x)\right)\right] +
    \frac{1}{2}\frac{\delta^2}{\delta\psi^2(x)}
      \left[-i \Gamma \psi^2(x)\right] \right\}P+\mbox{c.c.}
\end{equation}
As anticipated by the results for the single mode problem in 
section~\ref{sec:one-mode-prob}, the diffusion matrix
of this equation is non positive-definite and so
there is no straight forward mapping onto a single stochastic
differential equation~\cite{crispin2}.  Just as for the single mode case,
we move to a positive-$P$ representation and double the phase space 
with the mapping
\begin{align}
\psi(x,t) &\rightarrow \psi_1(x,t), \nonumber \\
\psi^+(x,t) &\rightarrow \psi_2(x,t),
\end{align}
where $\psi_1(x,t)$ and $\psi_2(x,t)$ are independent fields. As before,
we are guaranteed a positive-definite diffusion matrix and finally
obtain the pair of Ito stochastic equations
\begin{mathletters}
  \label{eq:Peqs}
\begin{eqnarray}
  \label{eq:Peqsa}
  i\partial_\tau \psi_1(x,\tau) &=& \calK
  \psi_1(x,\tau)+\Gamma\psi_2(x,\tau)
  \psi_1^2(x,\tau) 
   +\sqrt{i\Gamma} \psi_1(x,\tau) \eta_1(x,\tau),\\
  \label{eq:Peqsb}
  i\partial_\tau \psi_2(x,\tau) &=&-\calK
  \psi_2(x,\tau)-\Gamma\psi_1(x,\tau)
  \psi_2^2(x,\tau) +\sqrt{-i\Gamma}\psi_2(x,\tau)\eta_2(x,\tau),
\end{eqnarray}
\end{mathletters}
where the noise sources $\eta_1$ and $\eta_2$ are real, 
Gaussian and delta-correlated in time
and space: $\overline{\eta_i(x,\tau)\eta_j(x',\tau')}=
\delta_{ij}\delta(x-x')\delta(\tau-\tau')$.
Note that the mean number of atoms $\langle \hat{N}(\tau)\rangle = $
$\langle\intii dx\,\hpsid(x,\tau)\hpsi(x,\tau)\rangle=$ $\overline{\intii
  dx\,\psi_2(x,\tau)\psi_1(x,\tau)}$ is conserved in the ensemble
average but fluctuates during a single trajectory due to the complex
noise.
We remark that in practice, it is numerically more convenient to work with 
the complex conjugate equation to Eq.~(\ref{eq:Peqsb}). We have used
the form shown in order to make a clearer connection to 
Eqs.~(\ref{eq:onestoch}).

Although our derivation indicates that 
Eqs~(\ref{eq:Peqs}) allow one to calculate 
\emph{single-time normally-ordered} 
quantum field averages, it was shown by Drummond~\cite{drumthesis}
that they actually allow for \emph{multi-time 
time-normally-ordered averages} to be found. 
In general, an expression for an arbitrary 
time-normally-ordered average is obtained by replacing 
$\hat  \psi (x,t)$ by $ \psi _1(x,t)$, 
$\hat  \psi ^{\dag} (x,t)$ by $ \psi _2^(x,t)$, and 
the quantum averaging by the stochastic one: 
\begin{equation}\label{eq:PathInt}
\langle 
\bar T \hpsid(x,t) \cdots \hpsid(x',t')\ 
T \hpsi(x'',t'') \cdots \hpsi(x''',t''')
\rangle=
\overline{
\psi_2(x,t)\cdots\psi_2(x',t')
\psi_1(x'',t'')\cdots\psi_1(x''',t''')
}.
\end{equation}
Here, $T$ and $\bar T$ denote, respectively, direct 
and reverse time ordering of the field operators. 
The upper bar on the RHS of this relation denotes  
an averaging over the random trajectories $\{ \psi _1, \psi _2\}$, 
with the stochastic measure characterised constructively 
by Eqs~(\ref{eq:Peqs}). 
In other words, this is a {\em path integral\/} 
over trajectories $\{ \psi _1, \psi _2\}$, 
with Eqs~(\ref{eq:Peqs}) providing the measure over the  paths. 
In quantum field theory, 
quantum averages of the form in Eq.~(\ref{eq:PathInt}) appear 
in the well-known Keldysh diagram techniques~\cite{Keldysh}. 
They are a subset of the full set of Keldysh averages 
(which in general also contain $\hpsi$'s under the $\bar T$-ordering 
and  $\hpsid$'s under the $T$-ordering). 
For this subset of quantum averages, Eqs~(\ref{eq:Peqs}) 
are {\em fully equivalent\/} not just to the master equation,  
but {\em to the Heisenberg equations of motion for the 
 quantum field\/}, 
providing a constructive path-integral representations  
for these averages. 
(Moreover with external sources added to 
Eqs~(\ref{eq:Peqs}) they account for the full set of Keldysh 
averages, thus becoming fully equivalent to the 
Heisenberg equations, see~\cite{JPA3,Corresp}).
In the context of field theory, 
it is perhaps worth remarking on a helpful simplification
that results from the non-relativistic nature of the problem.
The density matrix at a time  $\tau=\tau_0$ 
may be mapped directly onto the  $P$, Wigner or positive-$P$
distributions at the same time. These are then  used as 
distributions for initial conditions in simulations.
In the general case, one must rather match the density matrix 
to the distributions at $\tau=-\infty$,
subject to the usual assumption of adiabatic turning on of the 
interaction~\cite{JPA3}.

The positive-$P$ representation is guaranteed to give exact results
for as long as the ensemble averages converge. However, we see below
in section~\ref{sec:Results} that the trajectories are prone to large
excursions from the mean which quickly cause the simulation to blow
up.  Such problems with the positive-$P$ representation are
well-known~\cite{smi89,gil97,wer97}, and occur especially in systems with
strong nonlinearity and weak (or vanishing) damping which is precisely
the situation in the present case of a trapped interacting Bose
condensate.  We believe this is the first case however, for which
divergent trajectories appear for realistic physical values. As such
it is indication of the likelihood of strongly non-classical behavior
outside the description of the GPE. It is important to realize that
the failure of the positive-$P$ in such cases is not indicative of
a genuine ``divergence'' in the sense of quantum electrodynamics but
merely represents a rapid (presumably exponential) growth in the
width of the distribution. So while in theory the distribution remains 
physically correct, in practice it becomes impossible to accurately sample
the whole distribution numerically.
In fact, this problem is strongly dependent on the parameter range chosen.
Drummond and Corney have successfully used the positive-$P$ representation
to simulate evaporative cooling of the condensate~\cite{dru98}.
The cooling problem is a case in which the positive-$P$ representation
can be expected to be more robust than in zero-temperature calculations for
two reasons: the atomic density  is much lower, at least initially,
and there is a considerable damping in the form of the RF field used
to remove the hotter atoms.

\subsection{Truncated Wigner representation}
\label{sec:Trunc-Wign-repr}
In the absence of exact stable methods, we are forced to consider
approximate simulation techniques. One approach that has proved
successful in optical problems is the ``truncated Wigner''
method~\cite{wer97}.  In a similar fashion to that of
section~\ref{sec:multi-mode-prob}, the master equation can be 
mapped onto a Fokker-Planck like equation for the Wigner
distribution~\cite{crispin2,wal94}, which returns symmetrized
expectation values as opposed to the normally ordered averages of the
$P$ representations.  That is to say, we define the Wigner
distribution by analogy with Eq.~(\ref{eq:Pdef}) as
\begin{equation}
  \label{eq:Wdef}
  W(\{\psi,\psi^*\},\tau)=\left.\rho^{\text{(sym.)}}(\{\hpsi,\hpsid\},\tau)
  \right |_{
    \begin{array}{l}
      \scriptstyle \hpsi\rightarrow \psi \\  
      \scriptstyle \hpsid\rightarrow \psi^*,
    \end{array}}
\end{equation}
where $\rho^{\text{(sym)}}$ denotes the density operator $\rho(\tau)$
symmetrically ordered with respect to the field-operators $\hpsi,\hpsid$.
(see Ref.~\cite{crispin2} for the connection of this definition
to more familiar expressions for the Wigner function).
Using the functional differentiation notation,
the operator correspondences take the form~\cite{crispin2},
\begin{align}
\hpsi \rho &\leftrightarrow
        \left(\psi+\frac{1}{2}\frac{\delta }{\delta \psi^*}\right)W, 
&&\hpsid \rho \leftrightarrow 
        \left(\psi^*-\frac{1}{2}\frac{\delta}{\delta \psi}\right)W, 
\nonumber \\
\rho  \hpsi &\leftrightarrow 
        \left(\psi-\frac{1}{2}\frac{\delta }{\delta \psi^*}\right)W, 
&&\rho \hpsid \leftrightarrow
        \left(\psi^*+\frac{1}{2}\frac{\delta}{\delta \psi}\right)W.
\end{align}
Using these relations in the master equation we find the Wigner function
evolution equation
\begin{equation}
\label{eq:wigpde}
\frac{\partial W(\psi,\psi^*)}{\partial \tau} = \intii \dx 
 i\left\{ \frac{\delta}{\delta \psi}
        \left[\calK \psi + \Gamma (|\psi|^2-1)\psi\right] -\frac{1}{4}
        \frac{\delta^3}{\delta^2 \psi\delta \psi^*} \psi
 \right\} W(\psi,\psi^*)+ c.c..
\end{equation}
In this case, there is no second derivative term---the 
diffusion matrix vanishes identically--- and the quantum noise acts via
third order derivatives as ``cubic noise''. 
Unfortunately, there is no simple mapping
from cubic noise to a stochastic representation~\cite{crispin2} 
and as we have discussed earlier, a direct integration of  
Eq.~(\ref{eq:wigpde}) is impractical.  
The simplest approximation is to truncate Eq.~(\ref{eq:wigpde}) at
second order so that we are left with a single deterministic equation
for the classical field $\psi_W$, which is just the standard
time-dependent GPE
\begin{equation}
  \label{eq:Weq}
  i\partial_\tau \psi_W(x,\tau) = \calK \psi_W(x,\tau)+\Gamma|\psi_W(x,\tau)|^2
\psi_W(x,\tau).
\end{equation}
Although this equation is completely deterministic, it is not the case
that we have discarded all effects of quantum noise. 
Noise is still included explicitly in the initial state
which is now represented as a distribution of functions $\psi_W(x,0)$.
(In fact, even in the positive-$P$ representation we would require a 
random distribution of starting functions unless the initial state was
a coherent state.)  We discuss the choice and representation of initial 
states in detail in section~\ref{sec:Initial-States}.  

In optical problems, it has been found that the Wigner
approach gives accurate results in the large photon number limit, when
it might be expected that the influence of the third order quantum noise
is small.  In section~\ref{sec:Results}, we test the Wigner
predictions against the positive-$P$ results for as long as the
latter are stable.  Using the Wigner distribution entails one
further limitation. Typically, the physically most interesting quantities
are time-ordered, normally-ordered averages, as provided directly by the $P$
representations~\cite{crispin2,wal94}. As the Wigner distribution
returns symmetrized moments and we do not know the unequal time
commutators for the field operators, we can not usually find multi-time
averages with the Wigner method. An exception is the two-time 
normally-ordered correlation function for coherent initial
states,
\begin{align}
\label{eq:twotime}
\langle \hpsid(x,\tau)  \hpsi(x',\tau')\rangle 
  &= \bra{\psigp(x,\tau)}\hpsid(x,\tau) 
  \hpsi(x',\tau')\ket{\psigp(x,\tau ) } \nonumber \\
  &=\psigp(x,\tau)^*\bra{\psigp(x,\tau)}
  \hpsi(x',\tau')\ket{\psigp(x,\tau)} \nonumber \\
  &=\psigp(x,\tau)^*\overline{\psi_W(x,\tau')},
\end{align}
which is thus reduced to a single time expectation value with no
ordering problems. Note that even for coherent initial states, higher-order
correlations such as 
$\smexpect{:\hpsid(\tau_1)\hpsi(\tau_1)\hpsid(\tau_2)\hpsi(\tau_2):}$ 
are unavailable.

\section{Initial States} \label{sec:Initial-States}
The question of suitable initial states for simulation is somewhat
involved.  Here we wish to use states that can be thought of as
a good representation of the ``ground state of the condensate'', 
in as much as this is possible in a symmetry broken picture.
We consider only zero temperature states here.  For $T=0$,
the simplest option is to choose an initial coherent state, that is,
precisely that state assumed at all times in a conventional
calculation with the GPE.  Our simulations then indicate how the
actual state evolves away from the coherent state.  To do so, we set
the mean field $\smexpect{\hpsi(x,0)}$ equal to the solution of the
time-independent GPE $\psigp(x)$ (the ``ground state wavefunction'')
and assume vacuum noise in all modes. For the normally-ordered
positive-$P$ representation, vacuum noise is obtained simply by the
choice $\psi_1(x,0)=\psi_2(x,0)=\psigp(x)$. In the
symmetrically-ordered Wigner representation, the noise must be
explicitly included in each mode of a suitable basis. Each trajectory
begins with a different field of the form
\begin{equation}
\psi_W(x,0) = \psigp(x) + \sum_{j=0}^N\frac{1}{2}\eta_j \phi_j(x)
\end{equation}
where $\eta_j$ is a complex random variable of zero mean with
$\overline{\eta_j \eta_k}=0$ and
$\overline{\eta_j^*\eta_k}=\delta_{jk}$. The sum is taken over $N$
modes of a complete basis $\{\phi_j(x)\}$ and
$N$ is taken sufficiently large that the results are independent of
$N$. Natural choices for the $\{\phi_k(x)\}$ are the discrete position
basis $\phi_j(x) = \delta(x-j\Delta x)/\sqrt{\Delta x}$ or the
harmonic oscillator basis. The
latter has the advantage that if $N$ is not too large, the modes do
not extend to the boundary of the simulation window and there is no
risk of noise artificially ``wrapping around'' the simulation. In
practice, we have seen no difference in results when using either of
these bases.

However, if one wishes to find a good approximation to the ground
state of the many body system, the coherent state is certainly not an
optimal choice.  Of course, as we are using a
symmetry-breaking approach, no state can be truly stationary---there
must always be a degree of phase diffusion associated with the number
superposition implied by the assumption of a non-zero mean field. Nevertheless,
there are states we might favour over the coherent state. Standard
applications of Bogoliubov theory at zero
temperature~\cite{fet72,gri96} approximate the second quantized
Hamiltonian by the diagonal expression
\begin{equation}
  \hH = K + \sum_{j>0} E_j \hbd_j\hb_j,
\end{equation}
where $K$ is a constant, $\hb_j$ is the annihilation operator for the
quasiparticle excitation of energy $E_j$ and the mean field satisfies
the GPE.  Further, the field operator may be written as
\begin{equation}
  \hpsi(x) = \sum_{j>0} [u_j(x) \hb_j-v_j^*(x)\hb_j^\dag],
\end{equation}
where the mode functions $u_j$ and $v_j$ are solutions to the
Bogoliubov-de Gennes eigenvalue equations~\cite{fet72,gri96}. Our second 
choice for the ground state is thus the vacuum in the Bogoliubov
representation:
\begin{equation}
\label{eq:plainbog}
\psi_W(x,0) = \psigp(x) + 
      \sum_{j=1}^N\frac{1}{2}(\eta_j u_j(x)-\eta_j^* v_j(x)).
\end{equation}
However, Lewenstein and You~\cite{lew96} have pointed out that in a
symmetry-breaking Bogoliubov method, the existence of a zero-energy
Goldstone mode requires the inclusion of an extra term in the
Hamiltonian involving the condensate ``momentum'' $P$ that accounts
for the phase diffusion of the mean field. In this case we have
\begin{equation}
  \label{eq:lewbog}
  \hH = K + \frac{\al}{2}P^2 + \sum_j E_j \hbd_j\hb_j,
\end{equation}
with $\al=N \partial \mu/\partial N$, $P = \intii \dx \psigp(x)
(\hdpsi+\hdpsid)$, and $\hdpsi = \hpsi-\smexpect{\hpsi}$.  This
Hamiltonian implies an infinite amplitude squeezing of the condensate
which is clearly unphysical. It has been shown elsewhere for one and
two mode models~\cite{dun98,ste98} that retaining cubic and quartic
terms in the Hamiltonian which are neglected in the Bogoliubov method
leads to a finite squeezing. Here, we do not perform a full treatment
of the effect of the higher order terms for the multi-mode system, but
take for our third ``ground state'', the lowest-energy  
variational state in which each mode in the Bogoliubov basis is
independently in a minimum uncertainty Gaussian state.

We also briefly remark that the choice of initial state is closely
tied to the manner of state preparation. In many instances, the
appropriate state need not be the ground state. In a recent experiment
at JILA~\cite{hal98}, a single condensate is subjected to a short
$\pi/2$ pulse creating a second condensate in a different internal
state. As the pulse length is shorter than the time required for
significant nonlinear dynamical effects to occur, the combined two
condensate system might be expected to exhibit binomial statistics. If
the trapping potential were arranged so that the two clouds did not
subsequently overlap, we could then model the evolution of one of the
two condensates assuming a number variance $(\Delta N)^2\approx N/2$. 
In fact, it may be checked using results in Ref.~\cite{lew96}, that the state 
in which all the Bogoliubov modes are in the vacuum has number statistics
for the condensate mode very close to $(\Delta N)^2\approx N/2$. 
For a slower transfer of population, nonlinear effects would play a
role and a more complex initial state would be
appropriate~\cite{leg98}.

As an example of a completely different initial condition,
the first-principles simulation of evaporative cooling using the positive-$P$
representation that was mentioned earlier~\cite{dru98},
begins essentially with a thermal state for the atom field.

\section{Results} \label{sec:Results}
\subsection{Numerical methods} \label{sec:Numerical-methods}
The parameters for our simulations are chosen to represent the
following system. We consider a condensate of $N=1000$ sodium atoms in
a cylindrical trap with $\lambda=\omega_z/\omega_r=0.025$ so that the
one-dimensional approximation is reasonable.  The radial frequency is
set at either $\omega_r/2\pi=800$~Hz (the ``strong trap'') or
$\omega_r/2\pi=200$~Hz (``weak trap''). Taking the scattering length
as $a=4.9$~nm we obtain for the nonlinear constant $\Gamma =
\Gamma_{\text{strong}} =0.084$  or
$\Gamma=\Gamma_{\text{weak}}=0.042$ . The initial state mean field
solutions were obtained by imaginary-time propagation of the GPE and
quasiparticle energies and mode functions found by standard
methods~\cite{fet72,lew96}.  The predictions of our simulations were
checked by ensuring that the results did not change when the time step
was decreased or the size of the spatial grid increased.  Simulations
in the truncated Wigner representation were performed with a standard
second-order split step method~\cite{wer97}.  Due to the large
nonlinearity of the system and consequently strong noise in the
positive-$P$ simulations, the standard Euler split-step algorithm was
not able to give results independent of time step even at a step size
of $d\tau=0.0005$.  Hence, we used a strongly-convergent semi-implicit
method~\cite{wer97} which gave reliable results with 256 spatial
points and a time step $d\tau=0.001$.

\subsection{Quantities of Interest}
We present our results in terms of three general quantities. To demonstrate
the ability to determine two-time correlation functions we would like to 
calculate the quantity
\begin{equation}
g^{(1)}(\tau,0) = \frac{\expect{\intii \dx \hpsid(x,0) \hpsi(x,\tau)}}{N}
\end{equation}
This is straight-forward in the positive-$P$ representation. However, as discussed in section~\ref{sec:Trunc-Wign-repr}, in the Wigner representation  we may only calculate
unequal-time normally ordered correlations for
coherent initial states, when from Eq.~(\ref{eq:twotime}) we have
\begin{equation}
g^{(1)}(\tau,0) =\frac{\overline{\intii \dx \psigp^*(x)\psi_W(x,\tau)}}{N}. 
\end{equation}
For other initial states, 
we can define a nominal ``condensate mode'' operator 
associated with the normalized solution to the GP equation,
\begin{equation}
\hagp(t) = \intii \dx \psigpbar(x) \hpsi(x,t),
\end{equation}
where $\psigpbar(x)=\psigp(x)/\sqrt{\intii \dx |\psigp(x)|^2}$. Its mean
value $\smexpect{\hagp(t)}$ still monitors the collapse of the wave function
but is strictly a one-time average and can be calculated in either
representation.We may also calculate the occupation of the condensate mode 
$\expect{\ngp}=\expect{\hagpd \hagp}$.

Finally, spatial correlations may be analyzed in terms of a spatial squeezing
spectrum. We define the localized amplitude quadrature operator 
\begin{equation}
\hX_\theta(x,\tau)= \psigpbar(x)\hpsid(x,\tau) e^{i\theta}
    +\psigpbar(x)\hpsi(x,\tau) e^{-i\theta}.
\end{equation}
Defining the Fourier transformed operator
\begin{equation}
  \hX_\theta(k,\tau) = 
  \frac{1}{\sqrt{2\pi}}\intii \dx e^{i k x}\hX_\theta(x,\tau),
\end{equation}
the squeezing spectrum is defined as the normally-ordered
expression~\cite{wal94,car87,dru87,wer95}
\begin{equation}
  \label{eq:squeeze}
  S_\theta(k,\tau) = 2\pi \expect{:\hX_\theta(-k) \hX_\theta(k):},
\end{equation}
which in the Wigner representation becomes
\begin{equation}
  S_\theta(k,\tau) = -1+2\pi \overline{\hX_\theta(-k) \hX_\theta(k)}.
\end{equation}
The angle $\theta$ for optimum squeezing is in general a function of $k$. 
Hence a useful quantity is the spectrum of ``best squeezing'' 
$S_{\text{max}}(k,\tau)$ which 
gives the largest possible squeezing at each wavenumber component $k$.

\subsection{Comparison of methods} \label{sec:Comparison-methods}
We first give some examples of calculations with the same parameters using
both the positive-$P$ and Wigner simulations. 
Figure~\ref{fig:strmeanval} shows the two time correlation function 
$g^{(1)}(\tau,0)$ for 1000 atoms in the strong trap configuration for a 
coherent initial state. Single
mode models~\cite{wri96,lew96} predict a Gaussian decay 
\begin{equation}
\label{eq:gauss}
g^{(1)}(\tau,0) = \exp[-\al^2\tau^2 (\Delta N)^2/(2N^2)]
\end{equation}
where $\al=N d\mu/dN$,
and $(\Delta N)^2=N$ is the variance in atom number of the initial state. This
model is indicated by the chain curve. The Wigner prediction (dotted) roughly
follows the Gaussian decay but shows slow oscillations about the single mode 
curve, and in particular exhibits a linear decay at short times. The Wigner 
method gives a stable result for arbitrary times. 
In the inset we show the short time behaviour, with the inclusion
of the positive-$P$ prediction in the solid line. This line stops at just
$\tau=0.3$ at which point unstable trajectories appeared. 
Also just visible are error bars on the positive-$P$ line denoting one 
standard deviation (OSD) uncertainties.  The dashed curves indicate the 
OSD errors for the Wigner calculation. 
The two methods clearly agree up to the point at which the unstable
trajectories arise. Note that the positive-$P$ error bars are very small
right up till that point, indicating the sudden rapidity with which
the distribution diverges.
In Fig.~\ref{fig:strnumber}, we show the occupation of the GP mode
$\smexpect{\ngp}$ as a function of time with the Wigner result
shown as the dotted line and the positive-$P$ result shown as the solid line. 
We see oscillations in the number with an amplitude of around 5~\% of the
starting population. Once more while the Wigner result is stable, the 
positive-$P$
simulations fail in a very short time. Note also that, the blow-up occurs
very rapidly---there is very good agreement until just before the fatal moment
with OSD errors for both methods being smaller than 
the thickness of the lines.

Thus for the trap parameters considered so far, the positive-$P$
representation is effectively useless. While instabilities of the
positive-$P$ are well-known this is perhaps the first occasion in
which they arise in experimentally accessible parameter ranges.  Given
that the positive-$P$ fails well before the completion of a single
oscillation in Fig.~\ref{fig:strnumber}, we might wonder how closely
the truncated Wigner results approximate the true dynamics. One
approach is to perform simulations at artificially low scattering
lengths for which the nonlinearity is less severe and the positive-$P$
simulations more robust. In fact the observation of Feshbach
resonances in an optically trapped $\text{Na}^{23}$
condensate~\cite{ino98} demonstrates that reduced scattering lengths
are now attainable in the presence of a sufficiently strong magnetic
field.  Figure~\ref{fig:vwknumber} shows the GP mode occupation as a
function of time for 1000 atoms in a coherent initial state, with the
reduced interaction $\Gamma = \Gamma_{\text{strong}}/10$. The line
styles have the same meaning as in Fig.~\ref{fig:strnumber}. The
positive-$P$ trajectories are now stable for much longer and it is
seen that both methods produce oscillations that are in agreement
within the error limits. Note that the error limits grow in time for
the positive-$P$ but remain approximately constant for the Wigner,
corresponding to the fact that no new noise is added after the initial
condition for the Wigner method. We can thus now have some confidence
that the Wigner calculations give results that are reasonably accurate
for relatively large condensates.

\subsection{Comparison of initial states}
We now examine the behavior exhibited by different initial states. 
As explained in section~\ref{sec:Initial-States}, we compare the standard
choice of coherent initial state with the vacuum state in the Bogoliubov 
representation and the Gaussian state with independent squeezing in each
Bogoliubov mode. Figure~\ref{fig:compinit} shows the mean amplitude in 
the GP mode $\smexpect{\hagp(\tau)}$ with Wigner results shown in the solid 
lines and the single mode estimates based on the initial number variance shown
in the chain lines.  The mean amplitude is apparently described relatively 
well by the single mode model. The differences between the curves
is largely accounted for by the difference in number variance in the
three cases, which had the values $(\Delta N)^2/N = $~1, 0.5 and~0.12 in the
coherent state, Bogoliubov vacuum and squeezed Bogoliubov vacuum cases
respectively.
Figures~\ref{fig:hstrsq}--\ref{fig:sqboghstrsq} give the spectra of best 
squeezing for the three initial states plotted as the function 
$\log[1+S_{\text{max}}(k,\tau)]$. For the coherent state in 
Fig.~\ref{fig:hstrsq}, there is of course initially no squeezing. 
For a short time, there is significant squeezing at low wavenumbers.
However, at large times the phase diffusion causes the long wavelength
fluctuations to grow without limit~\cite{lew96} and the squeezing is destroyed.
The other two initial states shown in Figs.~\ref{fig:boghstrsq} 
and~\ref{fig:boghstrsq} show similar trends at large $\tau$ but are
clearly different at early times when the statistics of the initial state
have not been swamped by the phase diffusion.  This suggests
that the squeezing spectrum may be a useful way of characterizing different
quantum states of the condensate.
Note that a single mode model could not predict different rates of change
for the squeezing at different wavenumbers.

\subsection{Negative scattering lengths}
Finally we briefly examine the dynamics for a single case with a negative
scattering length. In this case, the attraction between the atoms leads
to a high density at the center of the trap, and consequently the nonlinear
terms play a stronger role than in the positive scattering length case. 
To avoid the need of extremely fine spatial and temporal grids, we
therefore use the parameters of the weak trap, with the scattering length
set at $a_{\text{neg}} = -a_{\text{\small Na}}/10$ giving $\Gamma=-0.0042$. 
The two-time correlation function $g^{(1)}(\tau,0)$ and occupation 
$\smexpect{\ngp}$ display similar behavior to that seen earlier for the
positive scattering length. Here we concentrate on the squeezing spectrum 
which is shown in Fig.~\ref{fig:negsq}. This figure shows quite different
structure to the earlier squeezing spectra with strong anti-squeezing
at for wavenumbers near $k\approx3$,  corresponding to a length scale
of the condensate or ``soliton'' width.
In fact, this spectrum is very similar to the spectrum of best squeezing 
for a fiber soliton with a Kerr law nonlinearity~\cite{car87,dru87}. This
is not surprising. With a strong negative nonlinearity in a one-dimensional
trap, the condensate becomes strongly localized at the bottom of the trap.
The nonlinearity dominates over the trapping potential and the ground state
wavefunction is well-approximated by the fiber soliton expression 
$\psi(x)=\sqrt{N} \sech(\sqrt{N\Gamma}x)$, with a slight additional 
confinement due to the potential. Then as the propagation equations for
the two systems differ only by the inclusion of the potential for the
condensate, we can expect virtually identical spectra.

\section{Conclusion}
In this paper, we have applied phase space techniques for the
propagation of a complete quantum field to the problem of a
one-dimensional trapped Bose-Einstein condensate. As such systems are
highly nonlinear and weakly damped, the exact approach using the
positive-$P$ representation is useful only for short times compared to
the trap period and we are forced to use the approximate truncated
Wigner method. For parameter ranges in which both methods work, we
find agreement between the two.  The Wigner method is stable and
allows the calculation of one-time averages and certain conditional
multi-time averages over long periods. Dynamics may be calculated for
virtually any initial state with a reasonably well-localized Wigner
function.  

It is interesting to compare our approach here with another
set of tools for discussing quantum statistical properties of
condensates---the rapidly growing field of Quantum Kinetic Theory
(QKT), which has been developed in particular by Gardiner and Zoller
and their co-workers~\cite{QKT,QKTgrow}. 
In QKT, the system is divided into
two distinct parts---the ``condensate region'' consisting of the
condensate itself and a considerable number of the low-lying
excitations and the ``thermal region'' which is essentially everything
else and acts as a reservoir for the condensate region. One may then
obtain master equations for the condensate region of varying
complexity based on assumptions about the exchange of atoms between
the condensate and reservoir.  In our own approach, there is no
distinction at all into condensate and thermal atoms and thus no
approximations required in order to implement such a distinction. The
condensate itself plays no privileged role within the model and we
work simply with one complete quantum field.  The special properties
normally associated with condensates are manifested just as different
correlations of the quantum field.  
The stochastic method described here thus may also serve to provide
comparisons with the predictions of QKT from a rather different vantage
point.  Indeed, Drummond and Corney's simulations
of evaporative cooling using the positive-$P$ representation 
have produced~\cite{dru98} 
similar results to kinetic models of evaporative cooling~\cite{QKTgrow}.
Moreover, one might envisage a hybrid model in which the low-lying condensate
modes are treated using a stochastic approach while the upper modes
are reduced to a thermal reservoir using the techniques of QKT.

Finally, we point out some of the systems to which this theory could
be easily applied. As mentioned earlier, the coherence properties
of the output beams of atom lasers are certain to be of central importance
in the near future. Calculation of two-time correlations and
squeezing spectra for various laser designs is a natural application.
The phase diffusion between coupled condensates is also beginning to attract
interest and has currently only being studied theoretically within
the context of Bogoliubov theory~\cite{ruo98,law97}.

\acknowledgements
This research was supported by the Marsden Fund of 
the Royal Society of New Zealand, the University of Auckland Research Fund,
and NSF Grant PHY94-07194. MJS is grateful for the hospitality of Dr Weiping
Zhang during his stay at Macquarie University.

\begin{figure}[h]
  \begin{center}
        \caption{ \label{fig:strmeanval}
	Two-time correlation function $g^{(1)}(\tau)$ for coherent
          initial state in the strong trap. 
          Line styles are positive-$P$ (solid), Wigner
          (dotted), single-mode model (chain). Inset shows early times
          with 1 standard deviation errors for the Wigner method shown
          in dashed.}
   \end{center}
\end{figure}

\begin{figure}[h]
  \begin{center}
        \caption{ \label{fig:strnumber}
	Occupation number $\smexpect{\ngp}$ as a 
          function of time for strong trap.  Line styles are positive-$P$
          (solid) and Wigner (dotted).}
  \end{center}
\end{figure}

\begin{figure}[h]
  \begin{center}
        \caption{	\label{fig:vwknumber}
	Occupation number $\smexpect{\hat{n}_{\text{GP} }}$. 
        Line styles are positive-$P$ (solid) and Wigner (dotted).}
  \end{center}
\end{figure}

\begin{figure}[h]
  \begin{center}
        \caption{ \label{fig:compinit}
	Mean value $\smexpect{\hagp(t)}$. Wigner results are shown
        in solid lines, 1 mode models in chain. }
  \end{center}
\end{figure}

\begin{figure}[h]
  \begin{center}
        \caption{ \label{fig:hstrsq}
	Maximum squeezing spectrum plotted as 
          $\log[1+S_{\text{max}}]$ for weak trap parameters with coherent
                initial state.}
  \end{center}
\end{figure}

\begin{figure}[h]
  \begin{center}
        \caption{ \label{fig:boghstrsq}
	Maximum squeezing spectrum plotted as 
          $\log[1+S_{\text{max}}]$ for weak trap parameters with Bogoliubov vacuum
                initial state.}
  \end{center}
\end{figure}

\begin{figure}[h]
  \begin{center}
        \caption{ \label{fig:sqboghstrsq}
	Maximum squeezing spectrum plotted as 
          $\log[1+S_{\text{max}}]$ for weak trap parameters with
                squeezed Bogoliubov vacuum initial state.}
  \end{center}
\end{figure}

\begin{figure}[h]
  \begin{center}
        \caption{ \label{fig:negsq}
	Maximum squeezing spectrum $S_{\text{max}}$ for 
          system with attractive interactions: scattering length
          $a_{\text{neg}}=-0.049$~nm and $N=1000$ atoms.}
  \end{center}
\end{figure}

\end{document}